\documentclass[letterpaper,twocolumn,fleqn]{article} 

\usepackage{ist}
\usepackage{natbib}
\pdfoutput=1
\usepackage{dcolumn}
\usepackage{amsmath}
\usepackage{natbib}
\usepackage[T1, OT1]{fontenc}
\DeclareTextSymbolDefault{\dh}{T1}
\usepackage{hyperref}
\usepackage{color}
\usepackage{multirow}
\usepackage{listings}
\usepackage{algorithm}
\usepackage{algpseudocode}
\usepackage{graphicx}
\usepackage{caption}
\usepackage{subcaption}

\begin{document}

\title{StegoAppDB: a Steganography Apps Forensics Image Database}

\author{Jennifer Newman$^{+}$, Li Lin$^{+}$, Wenhao Chen$^{\dagger}$, Stephanie Reinders$^{+}$, Yangxiao Wang$^{\triangleright}$, Min Wu$^{\star}$, and Yong Guan$^{\dagger}$ \\
	$^{+}$Department of Mathematics, Iowa State University, Ames, Iowa, USA\\
	$^{\dagger}$Department of Electrical \& Computer Engineering, Iowa State University, Ames, Iowa, USA\\
	$^{\triangleright}$Department of Computer Science and Software Engineering, University of Washington Bothell, Bothell, Washington, USA\\
	$^{\star}$Department of Electrical \& Computer, University of Maryland, College Park, MD, USA}

\maketitle

\begin{abstract}
In this paper, we present a new reference dataset simulating digital evidence for image steganography. Steganography detection is a digital image forensic topic that is relatively unknown in practical forensics, although stego app use in the wild is on the rise. This paper introduces the first database consisting of mobile phone photographs and stego images produced from mobile stego apps, including a rich set of side information, offering simulated digital evidence. StegoAppDB, a steganography apps forensics image database, contains over 810,000 innocent and stego images using a minimum of 10 different phone models from 24 distinct devices, with detailed provenanced data comprising a wide range of ISO and exposure settings, EXIF data, message information, embedding rates, etc. We develop a camera app, Cameraw, specifically for data acquisition, with multiple images per scene, saving simultaneously in both DNG and high-quality JPEG formats. Stego images are created from these original images using selected mobile stego apps through a careful process of reverse engineering. StegoAppDB contains cover-stego image pairs including for apps that resize the stego dimensions. We retain the original devices and continue to enlarge the database, and encourage the image forensics community to use StegoAppDB. While designed for steganography, we discuss uses of this publicly available database to other digital image forensic topics.

\end{abstract}

\section{Introduction}

Creating effective tools for forensic practitioners requires that developers have access to standardized sets of data, as recognized by the National Academy of Sciences \cite{national2009strengthening}. As forensic processing of digital photographs becomes an increasingly important part of criminal investigations, the young field of digital image forensics must cultivate carefully designed and populated datasets. One area of digital image forensics is steganalysis, the analysis of a photograph for hidden content. Image steganography is the process to hide a \textit{message} or \textit{payload} in an \textit{innocent} image, producing a \textit{stego} image. See \cite{woodfordc} or \cite{kesslerg} for an introduction to steganography.

Stego apps on mobile devices are popular, easy to use, and stealthy. Table~\ref{t_apps_intro} displays the number of installs for 6 of over 100 apps that are available today. Development of techniques to discover steganography content where the image is "in the wild," that is, representative of what a practitioner may see while investigating a forensic case, is quite different from detection of a stego image in a controlled academic setting. We use ``in the wild" to describe unconstrained scenarios, involving many apps, many different source devices, and unknown processing to images such as photo editing. Access to simulated digital evidence images can permit benchmark of current steg detection software and advance improved solutions, as well as introduce academic researchers to unanticipated questions. Until StegoAppDB, no current image database provided data that is reflective of that found in mobile stego cases.

\vspace{0.3cm}
\begin{table}[h]
	\centering
	\caption{Real-world stego apps used for cover-stego image generation.}
	\label{t_apps_intro}
	\resizebox{\columnwidth}{!}{%
		\begin{tabular}{l|c|c|c|c}
			\hline
			\multicolumn{1}{c|}{\textbf{App Name}} & \textbf{Platform} & \textbf{\# Installs} & \textbf{\begin{tabular}[c]{@{}c@{}}Embedding\\ Domain\end{tabular}} & \textbf{\begin{tabular}[c]{@{}c@{}}Open\\ Source\end{tabular}} \\ \hline
			PixelKnot & Android & 100,000+ & JPEG & Yes \\
			Steganography\_M & Android & 10,000+ & Spatial & No \\
			Pocket Stego & Android & 1,000+ & Spatial & No \\
			MobiStego & Android & 1,000+ & Spatial & Yes \\
			Passlok Privacy & Android & 1,000+ & JPEG & Yes \\
			Pictograph & iOS & - & Spatial & Yes \\ \hline
		\end{tabular}%
	}
\end{table}

While image datasets have been used successfully for benchmarking academic steganalysis algorithms \cite{bas2011}, \cite{Gloe2010}, \cite{raise}, we identify some drawbacks for their use in benchmarking steganalysis tools on data closer to ``in the wild'' as encountered by forensic practitioners. For example, the commercial software StegoHunt \cite{stegohunt} advertises capabilities to analyze image data. However, we know of no publicly available datasets containing stego images on which to benchmark performance errors. Further, as the use of stego apps on mobile devices becomes more prevalent, large datasets containing examples of images from these sources will provide benchmarking capabilities for current and future software, allowing more realistic detection of steganography ``in the wild.''  Thus, creation of datasets that addresses some of these shortcomings is a welcome addition to the forensic community. Tellingly, the prevalence of steganography use occuring in forensic settings is unknown: there is no existing software designed for steg detection on suspect imags from a mobile device, nor any studies published detailing the population use rate of steganography, to the authors' knowledge.

If a forensic practitioner would like to test unknown images for steganograpy, she appears to have limited choices. First, three off-the-shelf software packages$-$StegoHunt \cite{stegohunt}, DC3 StegDetect \cite{dc3} and Provos' StegDetect \cite{provos}$-$cannot detect, with any reliability, stego images produced with recent stego algorithms \cite{chen2018forensic}. Second, if a forensic image analyst would like to develop or benchmark any new steg detection software beyond what these are designed to detect, there are no publicly available datasets (at least known to us).  Other steg detection approaches can be used, such as searching for evidence of auxiliary installation files on the computer \cite{zax2009faust}, which is proposed as field triage. 

We observe that when faced with a similar situation, the face recognition community turned to unconstrained datasets that challenged solutions beyond constrained datasets, such as driver’s licenses \cite{phillips2010frvt}, \cite{best2014}. In a similar manner, by using data sets that foster detection challenges with ``in the wild'' image data, the steganalysis community can pursue questions that are applicable to real-life scenarios, such as the transfer-learning problem of cover-source mismatch \cite{ker2013moving}. 

Motivated by this challenge, we propose that a database satisfy the following criteria for developing  and benchmarking solutions to practical steg detection.
\begin{enumerate}
	\item \textbf{Authentication}. Each image is provided with pedigree of origin, including camera device, meta data (including EXIF data), acquisition app, etc. 
	\item \textbf{Representation.} The data includes representatives of practical scenarios encountered in crime cases.
	\item \textbf{Evaluation.} The data is effective for evaluating and benchmarking standard algorithms including commercial software and academic algorithms, and allows reliable, reproducible, and measurable results that may be used in a court of law. 
	\item \textbf{Public access and free of copyright or privacy issues.} Communities require low-cost or free access to a standard data set without encountering copyright or privacy issues.
\end{enumerate}	

A review of several popular data sets used in digital image forensics research finds they have varying degrees of agreement with the four criteria, revealing a need for such data. Our goal is to provide an image database suitable to researchers in both crime lab and academic settings, allowing for performance evaluation on both academic steganalysis algorithms and commercial steganalysis software. Data in the database should allow simulation of "in the wild" digital evidence, as well as data appropriate for academic researchers to pursue problems related to real data created from real-world mobile stego apps. Should steganography evidence achieve the penultimate goal of being presented in a court case, performance evaluation on a standardized data set is in line with Daubert's requirement that scientific expert testimony be assessed for its evidentiary reliability \cite{stern2017statistical}.

In this paper we introduce a publicly available data set that agrees with much of 1-4 above. The database consists solely of images from mobile phones and mobile stego apps, representing data that certainly forensic practitioners see with increasing frequency. With over 800,000 stego and non-stego images created from stego apps on mobile phones, the data is offered in several file formats and with a wide variety of: scenes, exposure settings, embedding algorithms (software apps), embedding rates, devices (24) and models (10). A user-friendly web page provides detailed information on the content of the database and how to query and download. The data in StegoAppDB has extensive relational information for each data item. We continue to add to the database, and we invite suggestions to improve the contents or access. We anticipate access to a database with such varied and richly notated corpora will provide opportunities for motivated developers and researchers to create more practical solutions for steganography detection, and perhaps be useful for other digital image forensic purposes.

The remaining sections of the paper are as follows. In ``Related Works,'' we review popular data sets and current software for steg detection. In ``Creation of the Database,'' we discuss the acquisition procedure for original images and the generation of stego and other images for the database. In ``Descriptive Statistics, Substantiation and Evaluation,'' we give descriptive statistics and provide results of several experiments to substantiate our claims of the database$'$s investigatory nature. The section ``User Interface to Query for Data'' describes how the database can be queried. We conclude with remarks for forensic scientists from academic and practicing communities, including potential uses of our database in other areas of digital image forensics.

\section{Related Works}

While both academics and practitioners pursue forensic analysis of digital images, the two communities have very different goals. Academics seek innovative methods that advance the state-of-the-art in focused areas of conceptual performance; for steganalyisis, this can mean improving detection error, or introducing a new framework to improve performance, such as the relationship of embedding changes to syndrome coding \cite{crandall1998some}. At the other end of the spectrum, forensic practitioners expect their results to be interpreted in context of legal matters, and require outcomes that are validated by well-established and reproducible scientific procedures supporting quantitative analysis of uncertainty. Since our objective is to provide a set of data suitable to both communities, this section discusses issues from each community, including datasets, software and algorithms used to develop and benchmark steg detection.  Datasets used in the development of algorithms or software can, of course, influence the performance of the software. 

One typical job of an academic steganalyst is to create a new embedding algorithm (examples are WOW \cite{holub2012designing} or J-UNIWARD \cite{holub2014universal}), and then test its security (ability to be detected) using steganalysis techniques. Testing unknown images for hidden content with no prior information occurs only in stego challenges, the last of which completed in 2010 \cite{bas2011}, and another which is currently ongoing. As a standard practice, academic steganalyzers use a set of innocent images and create their own stego images using their code, and, for example, data from BOSSbase \cite{BOSSbase}, and more recently, for data-hungry convolutional neural networks, the BOWS-2 data set.  It is well-known that the peculiarities of a data set combined with features and machine learning classifier influences detection performance, e.g., see  \cite{toss}, where the authors verify different security performances based on compression or downsampling rates of RAW cover sources.  

While other image forensics are not the focus of our work, we note that two data sets created explicitly for image forensics -  the popular and publicly available data set for forgery detection RAISE \cite{raise} and the excellent Dresden database for camera identification \cite{Gloe2010} - have been used by researchers for steganalysis experiments. They are included below in our comparison with the four criteria. For descriptions of additional data sets, we refer the reader to the extensive review in \cite{raise}.

\begin{enumerate}
	\item BOSSbase \cite{bas2011} was created for an academic steganalysis competition in 2010 and by hindsight, introduced a clear example of the “cover-source mismatch problem” \cite{Bohme}. BOSSbase has been used successfully to evaluate the performance of hundreds of academic steganalysis algorithms. It contains 10,000 RAW images from 7 different still camera devices. However, it was not designed to benchmark commercial programs. The images have EXIF data, whose authentication, copyright and privacy statuses are unknown. The scenes reflect real-life scenarios, are collected in auto-exposure settings or half-auto settings, with the intent to produce high visual-quality images; thus, exposure settings do not cover more extreme lighting conditions.
	
	\item The RAISE database contains 8156 RAW images, constructed primarily for forgery detection \cite{raise}, in auto-exposure modes from three different still cameras over a period of several years. RAISE does not contain any JPEG images. It provides a challenging real-world data set They are authenticated, and the scene content reflects real-life scenarios. They are copyright-free, and most likely do not have privacy issues.
	
	\item The Dresden Image Database \cite{Gloe2010} was also created for the forensic community, with the main purpose for camera identification. It has almost 17,000 images, mostly in high-quality JPEG, from 73 still cameras devices representing 25 distinct models, with many different camera settings. However, the scene content is limited to a relatively small number of different scenes in order to replicate same scene/different camera scenarios for intense camera-specific forensic processing. Thus, it does not provide the wide range of scene content required for reliable stego detection. The data are authenticated and many images reflect real-life scenarios. They are copyright-free, and are taken with auto-exposure settings.
\end{enumerate}

Both RAISE and Dresden image data sets were not designed with steganography in mind, and so they lack stego images, naturally. Thus, their data cannot be used directly for evaluation of commercial steg detection software. While undoubtedly stego images could be produced and made available, currently this has not been done.

From an analysis of existing data sets and our requirements for creating a database to meet much of the four criteria, we propose our plan. The challenges were to meet as much of the four forensic criteria as possible, and generate hundreds of thousands of images.  Our data set consists of images produced from mobile phones and mobile apps, and retains provenanced side information for all images. The data is copyright-free and has no privacy issues, and is free and open to the public. To generate the images, we created a fast and efficient method using program analysis and reverse engineering. Each app has unique characteristics and required individual inspection using apk tools to generate the intermediate images and specific embedding rates. With these problems solved, StegoAppDB was populated. It is available online at \textit{https://forensicstats.org/stegoappdb/}~\cite{sad} and we encourage the forensic community to access, download and use the data, and contact us with any suggestions.

\section{Creation of the Database}

To populate our database, we use the four criteria as a guideline to develop appropriate data acquisition methods, data source choices, and auxiliary data authentication information. Our procedure was designed to collect a large amount of image data, to represent a reasonable number of different mobile phone cameras, and to include stego images from apps that are native to mobile devices. During the initial phase of data collection, we observed the well-known phenomenon that exposure settings of the images$-$related to image noise$-$impact the error rates of steganalysis algorithms \cite{lin2018impact}, \cite{lidomainadaptEI18}. Therefore, to make the image data representative, we collect the original photos with large diversities in exposure settings, which includes both ISO value and exposure time. To acquire such large amounts of photos from the phone's camera, we created our own research camera app that allows the photographer to collect 20 images automatically. Then, in order to create the large number of stego images from the stego apps on the phone, we reverse engineered each app individually using manual methods so that we could run the stego app directly on the phone that was cabled to the computer. This allowed us to produce stego images on the phone much faster than was possible by entering the same information on the app on the phone by a human. This section describes the process by which we create images that are put into our database.

\subsection{Collecting Original Images using Cameraw}

Exploiting the comprehensive range of features available for camera APIs on Android and iOS platforms, we create a camera app called ``Cameraw'' to capture images on our lab's smartphones \cite{cameraw}. Our main goal with Cameraw is to create a standardized process for the image acquisition procedure so that the app is simple to use, takes large amounts of photos quickly at acceptable visual fidelity but with varied exposure settings, and reduces the number of screen touches. By the press of one button, Cameraw automatically captures 20 images of one scene. After the ``capture'' button is pressed, the following steps ensue:

\begin{enumerate}
	\item The auto focus and auto exposure pre-capture sequence is triggered.
	\item After a short time, the focus is locked and exposure settings converge.
	\item Two auto exposure (AE) images are captured, one JPEG and one DNG, and 9 manual exposure settings are calculated using the AE values.
	\item The camera switches to manual exposure mode, and captures 9 pairs (one JPEG and one DNG) of additional images at the 9 manual settings, for a total of 20 images with 10 different exposure settings, within 15 seconds.
\end{enumerate}

Although the Android and iOS camera libraries provide auto exposure bracketing functionalities, we choose not to use them because they do not provide a wide enough range of ISO and exposure time values. Instead, we implement a customized bracketing method using the auto ISO and exposure time values that retain fairly good fidelity image quality. Let $i$ be the auto ISO, and let $e$ be the auto exposure time. We calculate 3 ISO values: $0.5*i$, $1.75*i$, $3.0*i$, and 3 exposure time values: $2.0*e$, $1.25*e$, $0.5*e$. From these values, we generate 9 distinct pairs of ISO and exposure time settings for capture in ``manual'' mode of the camera API. The other camera parameters are chosen by the built-in camera firmware. We lock all camera parameters during capture (except ISO and exposure time), to ensure all 20 photos have the same capture settings.

Cameraw is implemented with the camera2 API~\cite{camera2} in Android, and the AVFoundation framework~\cite{avfoundation} in iOS. We installed Cameraw on all devices, 10 Android phones and 14 iPhones. Several photographers were hired to capture a minimum of 100 indoor scenes using each device, resulting in total of 2412 scenes and 48,240 original images. We use the term ``original'' to describe those images that are captured using Cameraw on the mobile devices, with no further processing applied, including cropping. Our database contains other types of images that are created by processing methods, which we discuss in the next section.
%\vspace{-0.4cm}

\subsection{Generating Cover-Stego Images}

To facilitate benchmarking capabilities for steg detection of stego images generated from mobile stego apps, the database provides a large selection of app-generated stego images. We discovered that some apps embed a distinctive "signature" into the image, while others do not. We provide stego images for both types of apps.

\begin{figure}[t]
	\centering
	\includegraphics[width=\linewidth]{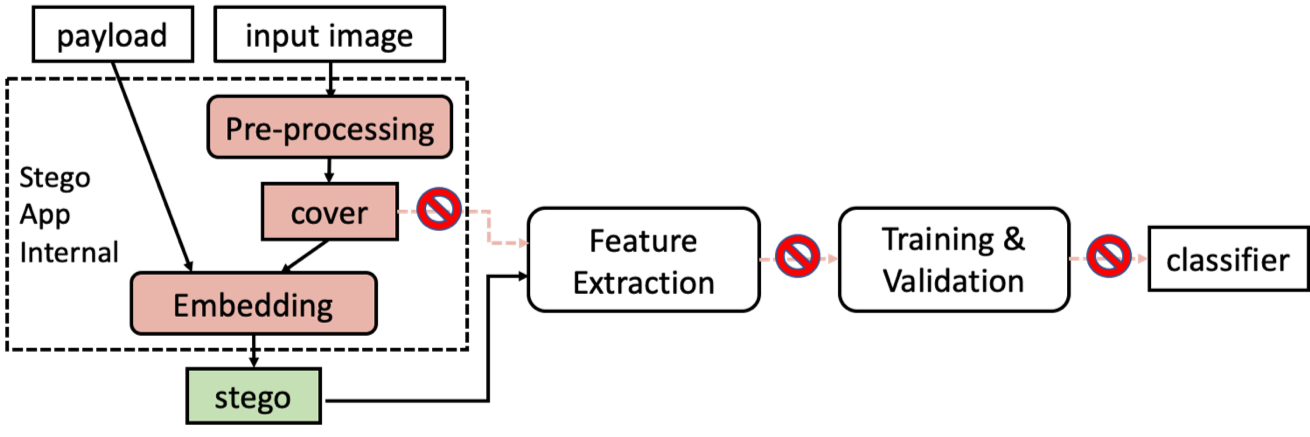}
	\caption{{\small{ Data and processing flow in a generic mobile stego app (left-hand side). On the right-hand side are steps to create a machine classifier, which cannot be completed for a stego app image due to unknown cover image.}}}
	\label{app_flow}
\end{figure}

We select six apps and create a coding environment on a computer, using a mobile phone connected to the computer, and batch-produce stego images with very specific embedding rates much quicker than possible by hand on the device. This process allowed us to save intermediate images where needed, specifically ``cover'' images. A cover image can be viewed as a 0\% embedded stego image, that is, a cover image is the image just prior to embedding and having the same pixel dimensions as the stego image. For many stego apps we observed, the image input to the app through the GUI user interface may undergo image resizing, among other alterations. Since academic steganalysis algorithms require cover-stego image pairs of the same dimension, machine learning classifiers in the traditional sense cannot be constructed from ``original'' image-stego image pairs as used by mobile stego apps. See Fig.~\ref{app_flow}, where the data flow for a generic mobile stego app is displayed on the left hand side. From the GUI of the stego app, the user inputs the payload, or message, as well as an input image, typically from the gallery or using the camera. The user may also enter a password as well. These data are passed to the internal code of the app, and any image processing such as resizing, or additional image creation, such as a cover image, can be obtained only through program analysis. Since the cover image is not known, feature extraction and subsequent machine learning development cannot be completed, as displayed on the right-hand side of Fig.~\ref{app_flow}. 

Cover images are a currently a necessity for machine learning algorithm development. This is not the case for certain stego apps that add an app-specific signature in the stego image that can be detected using other means. We next describe the process to generate cover and stego images for all apps.

%\vspace{-0.3cm}

\subsubsection{Stego Apps.}
To generate cover and stego images, we select six real-world stego apps: five Google Play Store apps and one Apple App Store app. See Table~\ref{t_apps_intro}. In the column ``Embedding Domain,'' we see that two apps embed payload in JPEG domain while the remaining embed in the spatial domain. The ``Open Source'' column indicates that source code for two apps is not publicly available. Inaccessible source code increases the difficulty of studying the apps' embedding process. For these two apps, we use reverse engineering tools to analyze the binary code and to delineate the steps that generate a stego image. We refer the interested reader to \cite{chenifip2018} for case studies of reverse engineering applied specifically to mobile stego apps.

We define the images in our database in the following way: An \textit{original image } is the image acquired by the mobile phone camera. An original image is used in many ways, such as an input image to a mobile stego app; or pieces are cropped to produce smaller-sized images to be used as cover images to embedding algorithms. A \textit{cover image} is the image that is paired to the stego image used for machine learning classifiers. A cover image has the same pixel dimensions (width x height) as the corresponding stego image, and can be viewed as a 0\% stego image. We define an\textit{ input image} to be the image that is selected by the user on the GUI of the stego app (on the phone), usually from the phone's gallery or by using the phone's camera. In academic steganalysis algorithms that use cover-stego image pairs, the input image as defined here is identical image to the cover image. In mobile stego apps, the input image is the photo selected by the user and subsequently processed by the app developer to create the final stego image. This input image can differ from the cover image due to pre-processing or resizing by the app prior to embedding. 

\subsubsection{Batch Image Generation.}
By manual code analysis of the apps, we determine how the embedding path is selected, which embedding method was used, pre-processing of input images and messages, etc. With knowledge of the data modification and processing steps, we create a script to batch generate cover$-$stego image pairs. Using source code modification and binary code instrumentation, we add two necessary functions to get complete side information for each stego image generated from a stego app:
\begin{enumerate}
	\item We save the intermediate cover image (see Fig.~\ref{app_flow}). This ensures that the database contains the corresponding cover image for each stego image, and is a critical step for stego apps that resize the input images prior to embedding. 
	\item We generate stego images with specific embedding rates. This is achieved by analyzing the stego app's embedding procedure, including how the payload is processed. To generate a stego image with a specific embedding rate, the modified app first calculates the cover image capacity and then calculates the necessary length of the embedded payload to achieve that rate. We correct for additional length added for auxiliary information.
\end{enumerate}

\begin{figure*}[t]
	\includegraphics[width=\textwidth]{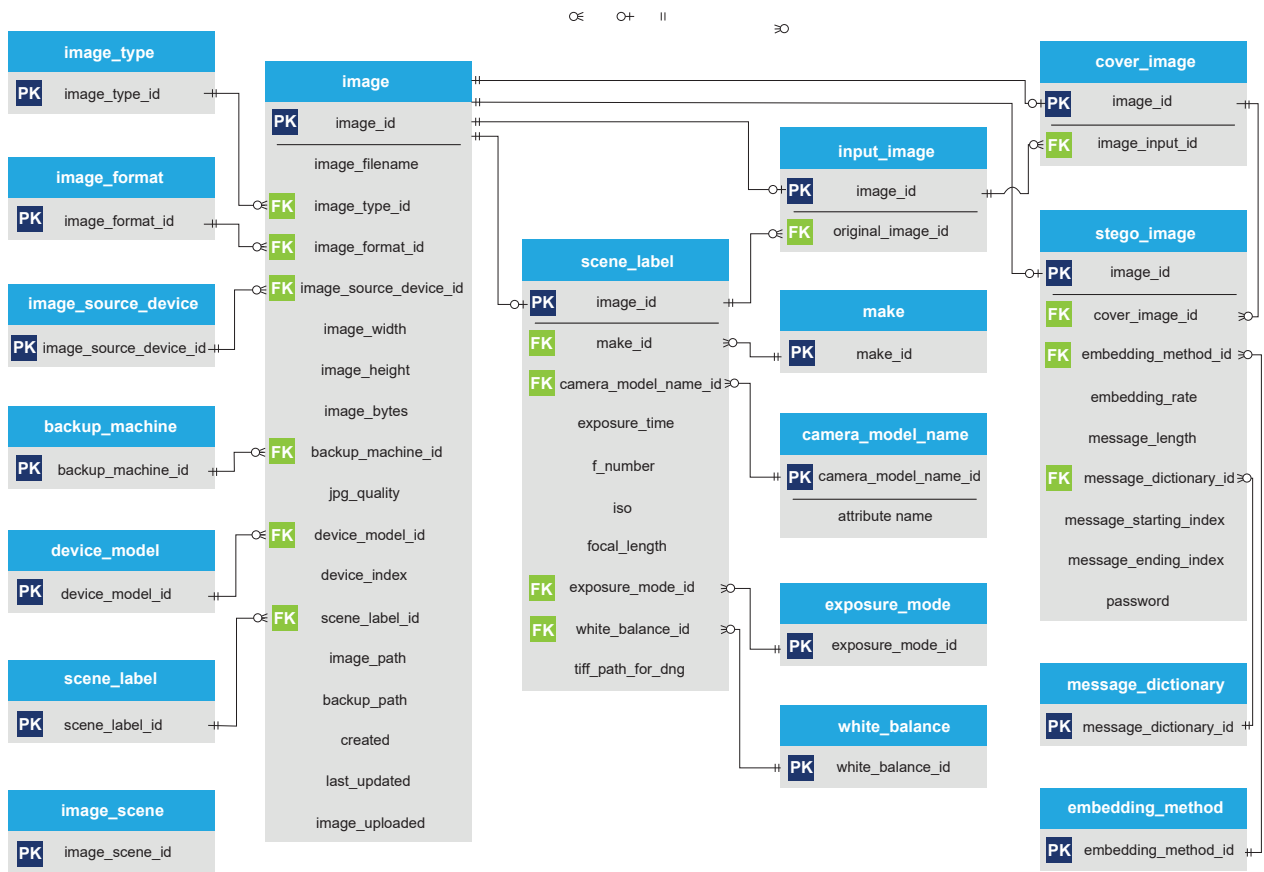}
	\caption{{{The entity-relation (ER) diagram for the database StegoAppDB.}}}
	\label{er_diagram}
\end{figure*}

As machine learning steganalysis classifiers prefer smaller images due to computational constraints, we process the original DNG and JPEG images to create (symmetrically) center-cropped grayscale PNG images (512x512). We use the term ``cropped'' to label these innocent PNG images. The PNG images are used as input images to the four spatial domain embedding apps to generate stego images. For the two frequency domain embedding apps, original-sized images are used as input images. Both DNG and JPEG images are used to generate \textit{PixelKnot} stegos, while only JPEG images are used to generate \textit{Passlok} stego images due to Passlok's inability to accept DNG files. All information is stored in the database for each stego image. We remark that all other variable being the same, the type of image used, DNG or JPEG, can affect the steg detection error rate \cite{toss}. The set of 10 JPEG and 10 DNG images from one scene (20 total images) differ only in file format and allow for experiments that involve compressed versus raw data comparisons.

The input messages embedded into the stego images are actual text messages selected from a set of dictionaries containing 634 Shakespeare plays~\cite{shakespeare}, ensuring variability by randomly selecting the start line for each text message from the dictionaries. The message dictionaries are provided with each download from StegoAppDB.

After this extensive coding process, we produce a large number of innocent images and corresponding stego images, at a variety of embedding rates, from a variety of phone models and devices, using a variety of currently-available mobile stego apps. The information used to generate each stego image is stored in data fields in StegoAppDB, including the associated cover image. Each cover image generated internally from an app is also stored in StegoAppDB, along with its information. The availability of this extensive side information for each image provides provenance and clarity about the generation process for each image.

The data collected and populated in StegoAppDB is arranged in a relational database.  The entity relation (ER) diagram is shown in Fig.~\ref{er_diagram}. This details the relation between cover images, input images, original images, and stego images, as well as the other information stored for each type of image.

\section{Descriptive Statistics, Substantiation, and Evaluation}

In this section, we present a quantitative summary of the data in StegoAppDB. We design experiments and give results that show how the use of different data in several current steg detection tools can substantiate strengths or weaknesses of the tool. We also select some academic machine-learning-based detection algorithms and evaluate their performance using the data from StegoAppDB.

\begin{table*}[t]
	\caption{Database summary: smartphone models, camera specifications, and number of images of different types.}
	\label{t_database}
	\resizebox{\textwidth}{!}{%
		\begin{tabular}{lcccccccc}
			\hline
			\multicolumn{1}{c}{Device Model} & \# Devices & \multicolumn{1}{l}{ISO Range} & \begin{tabular}[c]{@{}c@{}}Exposure Time\\ Range\end{tabular} & \# Scenes & \begin{tabular}[c]{@{}c@{}}\# Original\\ Images\end{tabular} & \begin{tabular}[c]{@{}c@{}}\# Cropped\\ Images\end{tabular} & \# Covers & \# Stegos \\ \hline
			Google Pixel 1 & 4 & 50 $\sim$ 3735 & 1/1258 $\sim$ 1/7 & 402 & 8040 & 8040 & 36180 & 180900 \\
			Google Pixel 2 & 2 & 50 $\sim$ 1708 & 1/9358 $\sim$ 1/9 & 201 & 4020 & 4020 & 18090 & 90450 \\
			Samsung Galaxy S8 & 2 & 56 $\sim$ 2097 & 1/2643 $\sim$ 1/12 & 208 & 4160 & 4160 & 18720 & 93600 \\
			One Plus 5 & 2 & 100 $\sim$ 3200 & 1/2777 $\sim$ 1/8 & 201 & 4020 & 4020 & 18090 & 90450 \\
			iPhone 6s & 2 & 25 $\sim$ 1000 & 1/60 $\sim$ 1/3 & 200 & 4000 & 4000 & 4000 & 20000 \\
			iPhone 6s Plus & 2 & 25 $\sim$ 1250 & 1/67 $\sim$ 1/3 & 200 & 4000 & 4000 & 4000 & 20000 \\
			iPhone 7 & 4 & 20 $\sim$ 1250 & 1/67 $\sim$ 1/3 & 400 & 8000 & 8000 & 8000 & 40000 \\
			iPhone 7 Plus & 2 & 20 $\sim$ 1000 & 1/60 $\sim$ 1/5 & 200 & 4000 & 4000 & 4000 & 20000 \\
			iPhone 8 & 2 & 20 $\sim$ 800 & 1/60 $\sim$ 1/3 & 200 & 4000 & 4000 & 4000 & 20000 \\
			iPhone X & 2 & 20 $\sim$ 800 & 1/62 $\sim$ 1/3 & 200 & 4000 & 4000 & 4000 & 20000 \\ \hline
			\multicolumn{1}{c}{total} & 24 & 20 $\sim$ 3735 & 1/9358 $\sim$ 1/3 & 2412 & 48240 & 48240 & 119080 & 595400 \\ \hline
			\multicolumn{1}{c}{total images} &  & \multicolumn{1}{l}{} & \multicolumn{1}{l}{} &  &  &  &  & 810960 \\ \hline
		\end{tabular}
	}
\end{table*}

\subsection{Descriptive Statistics of the Database Contents}

The images in StegoAppDB comprise original images, grayscale PNG images, and stego images and their corresponding cover images. Table~\ref{t_database} displays a summary of camera models and image data for the 24 devices. Each smartphone device acquired at least 100 indoor scenes of images, where one scene has 20 images. 

\begin{figure}[h]
	\centering
	\includegraphics[width=\linewidth]{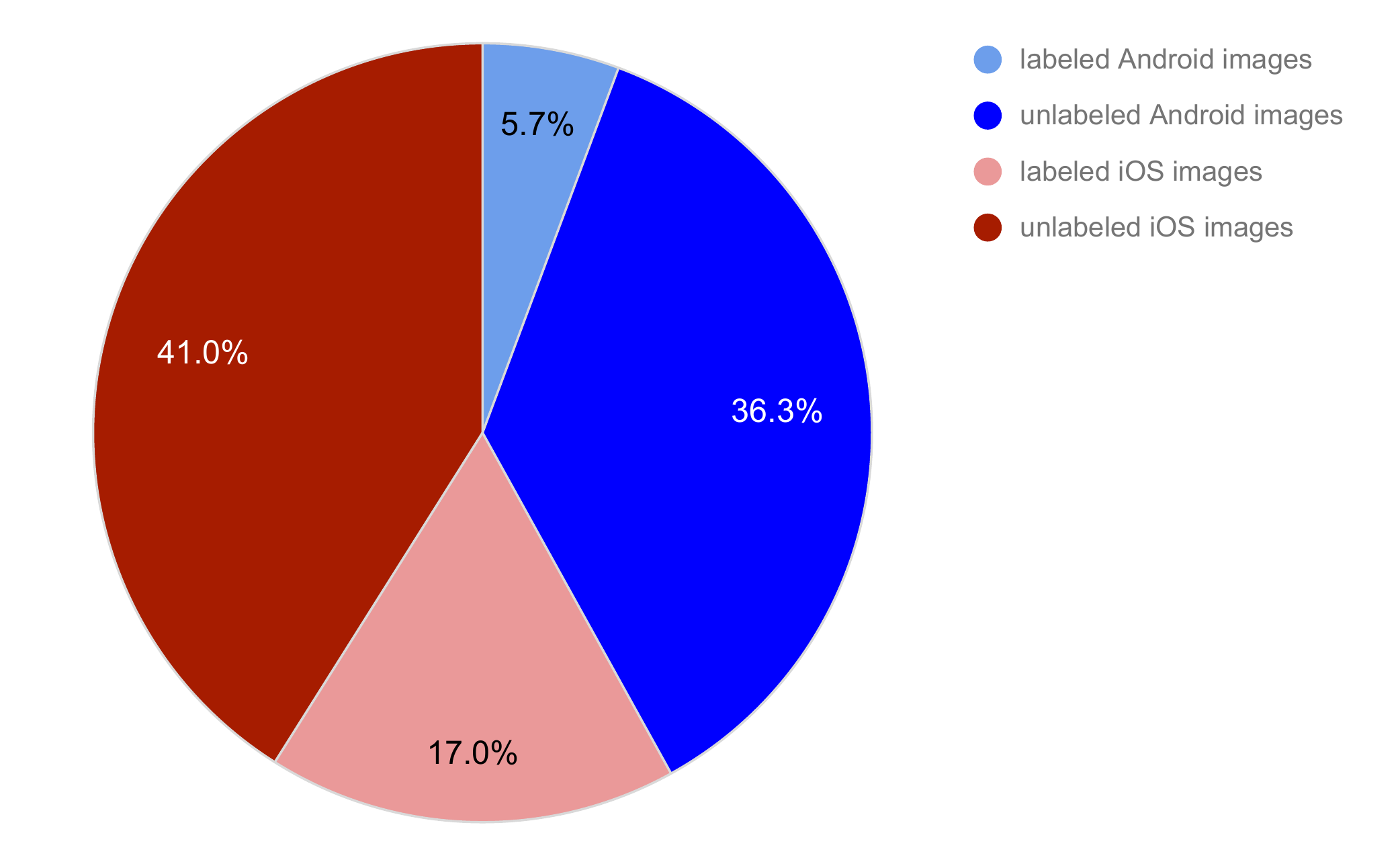}
	\caption{{Percentages of labeled vs. unlabeled images, identified by operating system.}}
	\label{fig_labels}
\end{figure}

\begin{figure}[h]
	\centering
	\includegraphics[width=\linewidth]{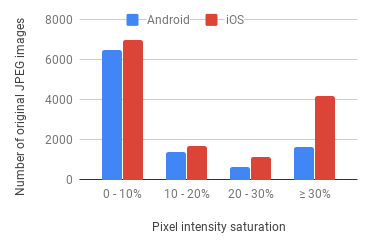}
	\caption{{ Distribution of saturation of intensity values}.}
	\label{fig_saturation}
\end{figure}

During photo acquisition, the photographer assigns a scene one of ten labels, or is unlabeled: \textit{books}, \textit{apple}, \textit{orange}, \textit{chair}, \textit{stairs}, \textit{backpack}, \textit{clock}, \textit{keyboard}, \textit{bottle}, and \textit{keys}.Figure~\ref{fig_labels} shows the number of labeled and unlabeled original images. Out of the 48240 original images, 10940 have labels.

In Fig.~\ref{fig_saturation}, we give a distribution of the number of images having saturated pixel intensity values. For an original JPEG image, we compute the proportion of intensity values across all three RBG planes that are below 5 and above 250. Each image falls into one of the four categories as given in Fig.~\ref{fig_saturation}, which shows their distribution across all original images.

\begin{figure}[]
	\centering
	\includegraphics[width=\linewidth]{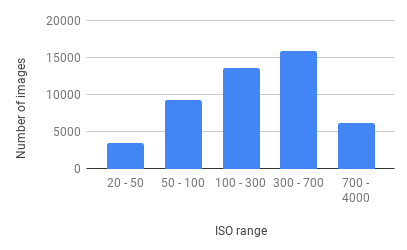}
	\caption{Distribution of ISO values of original images.}
	\label{fig_distr_iso}
\end{figure}

\begin{figure}[]
	\centering
	\includegraphics[width=\linewidth]{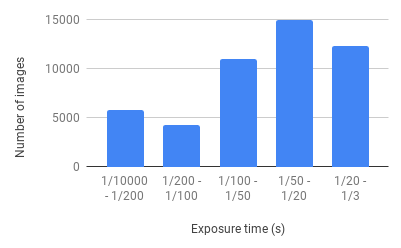}
	\caption{Distribution of exposure times of original images.}
	\label{fig_distr_expos}
\end{figure}

As the exposure settings varied for 18 of the 20 original images in a scene set, these images cover a wide range of ISO and exposure time values. Figures~\ref{fig_distr_iso} and ~\ref{fig_distr_expos} show the distribution of original images at different ranges of exposure settings.

As shown in Table~\ref{t_stegos_by_apps}, our database contains a total of 810,960, of which 595400 are stego and 119080 are cover. For each cover image, five stego images are generated at the embedding rates $0.05$, $0.10$, $0.15$, $0.20$, and $0.25$. Parameters relevant to the embedding process such as embedding rate, change rate, input message, password, etc., for each stego image are included in the csv files downloaded with each set of images.

\vspace{0.3cm}
\begin{table}[h]
	\centering
	\caption{Number of cover and stego images by Stego Apps.}
	\label{t_stegos_by_apps}
	\begin{tabular}{lccc}
		\hline
		\multicolumn{1}{c}{\textbf{Stego App}} & \textbf{\begin{tabular}[c]{@{}c@{}}Operating\\ System\end{tabular}} & \textbf{\begin{tabular}[c]{@{}c@{}}\# cover\\ images\end{tabular}} & \textbf{\begin{tabular}[c]{@{}c@{}}\# stego\\ images\end{tabular}} \\ \hline
		PixelKnot & Android & 20240 & 101200 \\
		Steganography (by Meznik) & Android & 20240 & 101200 \\
		Pocket Stego & Android & 20240 & 101200 \\
		MobiStego & Android & 20240 & 101200 \\
		Passlok Privacy & Android & 10120 & 50600 \\
		Pictograph & iOS & 28000 & 140000 \\ \hline
		total &  & 119080 & 595400 \\ \hline
		\newline
	\end{tabular}
\end{table}

Commercial or free programs assert capabilities of identifying steganography in some files or work environments. For example, the embedding algorithm F5 \cite{F5} has been widely used to hide messages in JPEG images, and three software programs $-$  StegoHunt\cite{stegohunt}, DC3-StegoDetect\cite{dc3} and Provos$-$StegoDetect \cite{provos} $-$ claim to detect stego images created by the standard F5 algorithm. To verify this, we randomly select 2000 original JPEG images from our database, implement the standard F5 algorithm to create 2000 stego images with 10\% embedding rate, and present all 2000 cover-stego pairs to the programs. The result is provided in Table \ref{evaluation_F5}.

\vspace{0.3cm}
\begin{table}[h]
	\centering
	\caption{Error of detection on images generated by standard F5}
	\label{evaluation_F5}
	\resizebox{\columnwidth}{!}{%
		\begin{tabular}{crrr}
			
			\hline
			\textbf{Error Type } & \textbf{Stego Hunt  } & \textbf{  DC3$-$StegDetect } & \textbf{ Provos$-$StegDetect} \\ \hline
			False Alarm         & 0\%                 & 0\%                     & 24.6\%                     \\
			Misdetection        & 5.2\%               & 0\%                     & 47.4\%                     \\
			Avg. Error          & 2.6\%               & 0\%                     & 36.0\%                    
		\end{tabular}
	}
\end{table}

As we can see from the Table~\ref{evaluation_F5}, both Stego Hunt and DC3$-$StegDetect have very good performance in detecting the stego images generated by the standard F5, while the Provos$-$StegDetect has an unacceptable error rate in our experiment. 

However, images generated by stego apps are very different from the images created by standard academic embedding algorithms. In this case, we use the app PixelKnot as an example, since it also implements the F5 steganography algorithm with some minor modifications. Again, we perform a different random selection of 2000 cover-stego pairs from our database that are created by PixelKnot, and present them to the three software programs. The result is presented in Table~\ref{Evaluation_PixelKnot}.

\vspace{0.3cm}
\begin{table}[h]
	\centering
	\caption{Error of detection on images generated by PixelKnot}
	\label{Evaluation_PixelKnot}
	\resizebox{\columnwidth}{!}{%
		\begin{tabular}{crrr}
			\hline
			\textbf{Error Type} & \textbf{  Stego Hunt} & \textbf{  DC3$-$StegDetect} & \textbf{ Provos$-$StegDetect} \\ \hline
			False Alarm         & 0\%                 & 0\%                     & 24.6\%                     \\
			Misdetection        & 100\%               & 100\%                   & 75.4\%                     \\
			Avg. Error          & 50\%                & 50\%                    & 50\%                      
		\end{tabular}
	}
\end{table}

As we can see from the Table~\ref{Evaluation_PixelKnot}, all three programs fail to detect the stego images created by PixelKnot. The developers' code modification of F5 implemented in their version of PixelKnot includes an omission of a signature of the F5 algorithm when writing the stego output images. The signature is found in stego images output from the standard F5 code. This shows that the rise of mobile stego apps can bring new challenges to the forensic image analyst. Indeed, if a digital forensic tool fails in a particular scenario, it can be argued that those results should be used to correct the tool weakness \cite{beckett2007digital}. With our comprehensive stego database, we are eager to work with research teams or companies that have strong interest in this new challenge and opportunity $-$ detecting stego images from mobile stego apps.

\subsection{Evaluation of Machine Learning Detection Algorithms}

It is well-known that machine learning is a very powerful tool in detecting stego images created by academic steganography methods. To show that it is also very effective in detecting stego images by apps, we design the following experiment.

Here, we target classification of stego images created by four Android apps:  PixelKnot, Steganography, Pocket Stego, and Passlok Privacy. To that end, we select original JPEG images from four different devices: one Google Pixel 1, one Google Pixel 2, one Samsung Galaxy S8, and one OnePlus 5, and all corresponding stego images. With embedding rate fixed at 10\%, for each device and each app, we have 1000 cover-stego pairs of images from the database, in which 500 pairs are used for training, and the remaining 500 pairs are used for testing.

For the machine learning methods, we implement the CC-JRM \cite{cc-jrm} for feature extraction on JPEG images (stego images created by PixelKnot and Passlok) and SRM \cite{srm} for feature extraction on PNG images (stego images created  Steganography and Pocket Stego). The FLD ensemble classifier \cite{ensemble}, which is essentially a random forest method, performs the classification. The classification accuracy is present in Table~\ref{Evaluation_four_four}.

\vspace{0.3cm}
\begin{table}[h]
	\caption{Classification accuracy of detecting cover-stego pairs by ML algorithms}
	\label{Evaluation_four_four}
	\resizebox{\columnwidth}{!}{%
		\begin{tabular}{lrrrrr}
			\hline
			\textbf{Apps}            & \multicolumn{1}{l}{\textbf{Pixel 1}} & \multicolumn{1}{l}{\textbf{  Pixel2}} & \multicolumn{1}{l}{\textbf{ Samsung S8}} & \multicolumn{1}{l}{\textbf{ One Plus 5}} & \multicolumn{1}{l}{\textbf{ Mix of four devices}} \\ \hline
			\textbf{PixelKnot}       & 97.5\%                               & 97.6\%                              & 97.6\%                                  & 98.3\%                                  & 99.0\%                                           \\
			\textbf{Steganography}   & 98.0\%                               & 97.8\%                              & 99.4\%                                  & 97.7\%                                  & 98.6\%                                           \\
			\textbf{Pocket Stego}    & 96.8\%                               & 97.3\%                              & 99.5\%                                  & 98.3\%                                  & 98.4\%                                           \\
			\textbf{Passlok Privacy} & 99.0\%                               & 97.1\%                              & 98.3\%                                  & 98.3\%                                  & 98.6\%                                          
		\end{tabular}
	}
\end{table}

As we can see in Table~\ref{Evaluation_four_four}, machine-learning-based detection algorithms have very impressive performance in detecting stego images, provided that we know which app create them. However, we point out that, in the above experiment, we only use JPEG images as the original source. In the case where RAW images are used as the original source to generate the stego images, the accuracy drops significantly \cite{chentackling2018}. Embedding rate can also affect the detection rate, and in realistic scenarios for small messages, the effective embedding rate can be less than 3\%, making detection even harder. In this experiment, only a few machine learning algorithms are testified. We welcome others to develop better detection algorithms for stego images from apps and test using our database.

\section{User Interface to Query for Data}

\begin{figure*}[t]
	\centering
	\includegraphics[width=\linewidth]{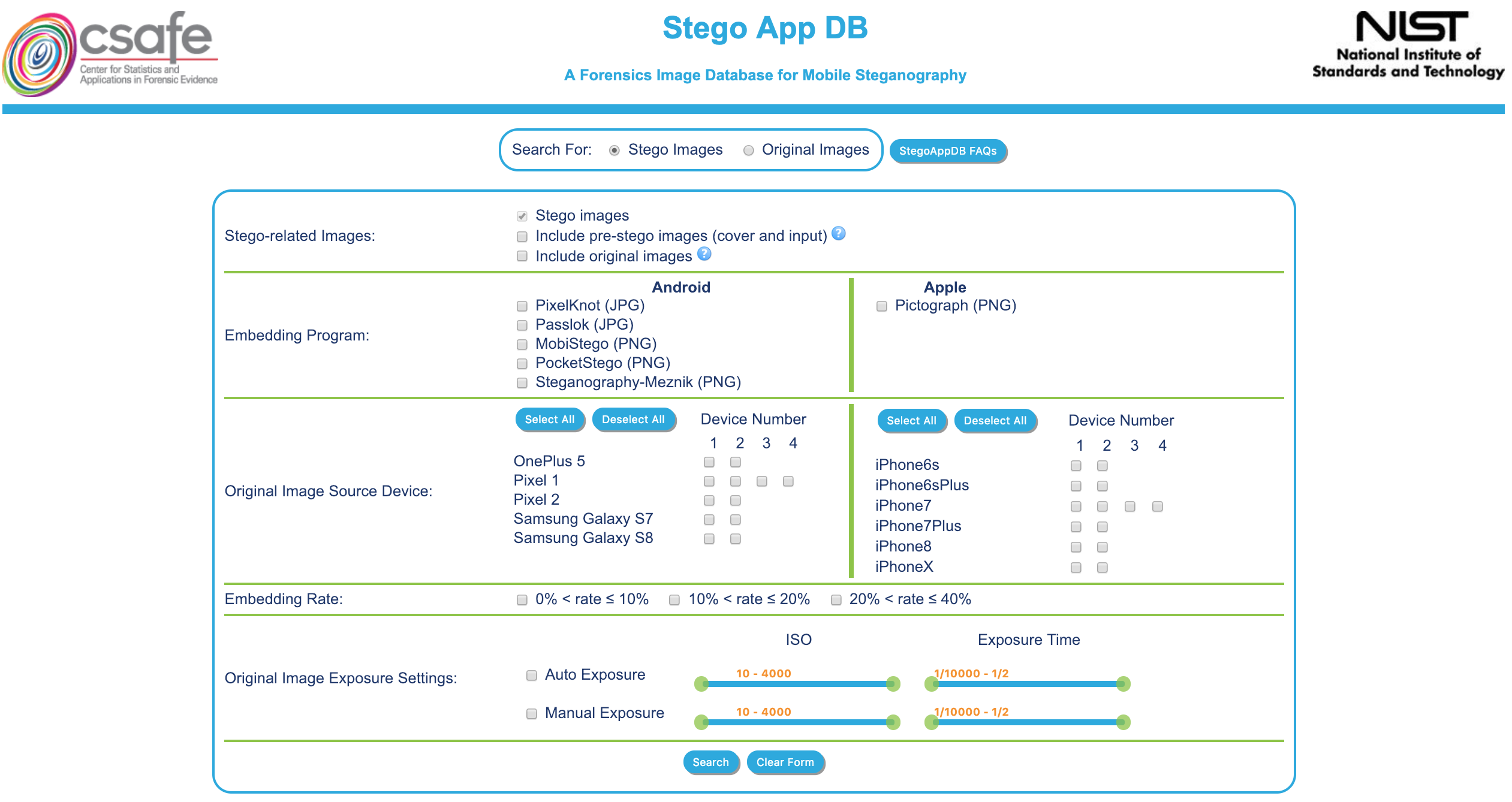}
	\caption{Search options for stego images in StegoAppDB.}
	\label{stego_web}
\end{figure*}

The design of the webpage for public access was created to facilitate as simple a process as possible for data queries, given the richness of the data. The database can be accessed through a link from CSAFE's webpage \cite{sad}. On the database's main webpage, we give descriptions of the information the user is able to query, listing the query parameters to search on a wide range of image characteristics.

There are two main types of data: 

\begin{itemize}
	\item ``Original'' images, which are the images captured using Cameraw on mobile devices, with native pixel dimensions as determined by the camera;
	\item ``Stego'' images, which are images embedded with messages and produced using the stego apps on mobile devices.
\end{itemize}

Original images or portions thereof may be used for academic steganalysis algorithms, forgery, or camera identification problems. Original images can be downloaded and portions cropped out by the user, or downsized to produce images appropriate for academic embedding algorithms. When querying for stego images, options are available to download other images that are associated with the stego images, such as cover images or input images. Downloading a cover image that generated the stego image would be useful in constructing a steganalysis machine learning classifier where cover images are required for training. (Recall that some apps downsize the input image prior to embedding, and the downsized image is not available to the app user.)

When a set of images are downloaded, a zip file is provided. The contents of the zipped file are: folders in which the images reside, one folder for each type of image (stego, cover, input, and originals); a csv file for each image type downloaded; a README text file that describes the search parameters for that particular search query; and a zip file containing the message dictionaries.

To find ``stego'' images, a user can query with parameters related to both the stego images and the corresponding source (original) images. There are five filter options for this query:
\begin{itemize}
	\item \textbf{Input/Cover Images.} A user can choose whether or not to download the input and cover images associated with the stego images.
	\item \textbf{Stego Program.} Select from six different stego apps.
	\item \textbf{Embedding Rate.} Select from different ranges of embedding rates.
	\item \textbf{Source Image Device Model.} Select to find stego images based on the device models from which the corresponding original images are collected.
	\item \textbf{Source Image Exposure Settings.} Select to filter stego images based on the exposure settings of the corresponding original images.
\end{itemize}

See Fig.~\ref{stego_web} for a picture of the stego search options.

The search for original images is similar to stegos. To find ``original'' images, a user can query the database by using the following filters.
\begin{itemize}
	\item \textbf{Device Model.} Select from any of the 10 device models and 24 devices.
	\item \textbf{Image Format.} Select from JPEG and JPEG quality, or DNG.
	\item \textbf{Scene Content.} Select from 10 indoor scene labels, or select ``unlabeled.''
	\item \textbf{Exposure Settings.} Select from auto exposure and specific ranges of ISO and exposure times.
\end{itemize}

With each image identified in a query, all meta information and side information are also provided. For original images, this includes EXIF data, device and label information. For stego images, in addition to EXIF data, all embedding characteristics including stego app, input message, password, etc. are provided. 

\section{Conclusion}
In this paper, we announce the new database StegoAppDB, the first data set consisting of hundred of thousands of images from mobile phones. The database contains images from 24 different devices of 10 different models, and includes stego images generated by stego apps from the phones. It has a variety of different types of data that make it amenable not only for steganography, but for forgery and camera identification. Each image has a rich set of annotated side information, free from copyright and privacy issues, and is publicly available. 

Other digital image forensic topics may find our database suitable. Recently it was shown that additional large steganography datasets can help solve other digital image forensic including detection of multiple image manipulations using Convolutional Neural Networks. Here, learned parameters from a CNN trained for steganalysis were "transferred" to a new CNN, the latter of which was trained successfully on a small amount of data from a new database to detect median filtering, JPEG compression, etc. Having more data sets available to the digital image forensics community can be a valuable resource.

We invite the forensic image analysis community to use StegoAppDB and offer suggestions for improvements and future content.
	
\section{Acknowledgement}
We are grateful to the following undergraduate students for helping us acquire images for our database: Yiqiu Qian; Joseph Bingham; Chase Webb; and Mingming Yue. This work was partially funded by the Center for Statistics and Applications in Forensic Evidence (CSAFE) through Cooperative Agreement \#70NANB15H176 between NIST and Iowa State University, which includes activities carried out at Carnegie Mellon University, University of California Irvine, and University of Virginia.

\end{document}